\documentstyle[twocolumn,aps,epsf]{revtex}
\begin{document}
\title{Decoherence and the Quantum Zeno Effect
\thanks{Based on a poster presented at the Workshop on Advanced Laser
Spectroscopy, IIT Kanpur, India, 25-28 February 1995}}
\author{Anu Venugopalan and R. Ghosh}
\address{School of Physical Schiences, Jawaharlal Nehru University,
New Delhi-110067, INDIA.}
\maketitle

The quantum Zeno effect (QZE) is the inhibition of transition between
quantum states as a result of frequent or continuous observations on the
state [1].  Ghirardi et al [2] have shown through general arguments based
on time-energy uncertainty relations that it is extremely difficult to
observe this effect in the case of spontaneous decay. Recently Nakazato
et al [3] have also analysed the QZE in the case of neutron spins and shown
that the limit of `continous measurements' is unphysical. QZE was observed
experimentally by Etano et al [4] in an rf transition between two
$^{3}Be^{+}$ ground state hyperfine levels 1 and 2 (Fig. 1).
\epsfysize=0.9in
\centerline{\epsfbox{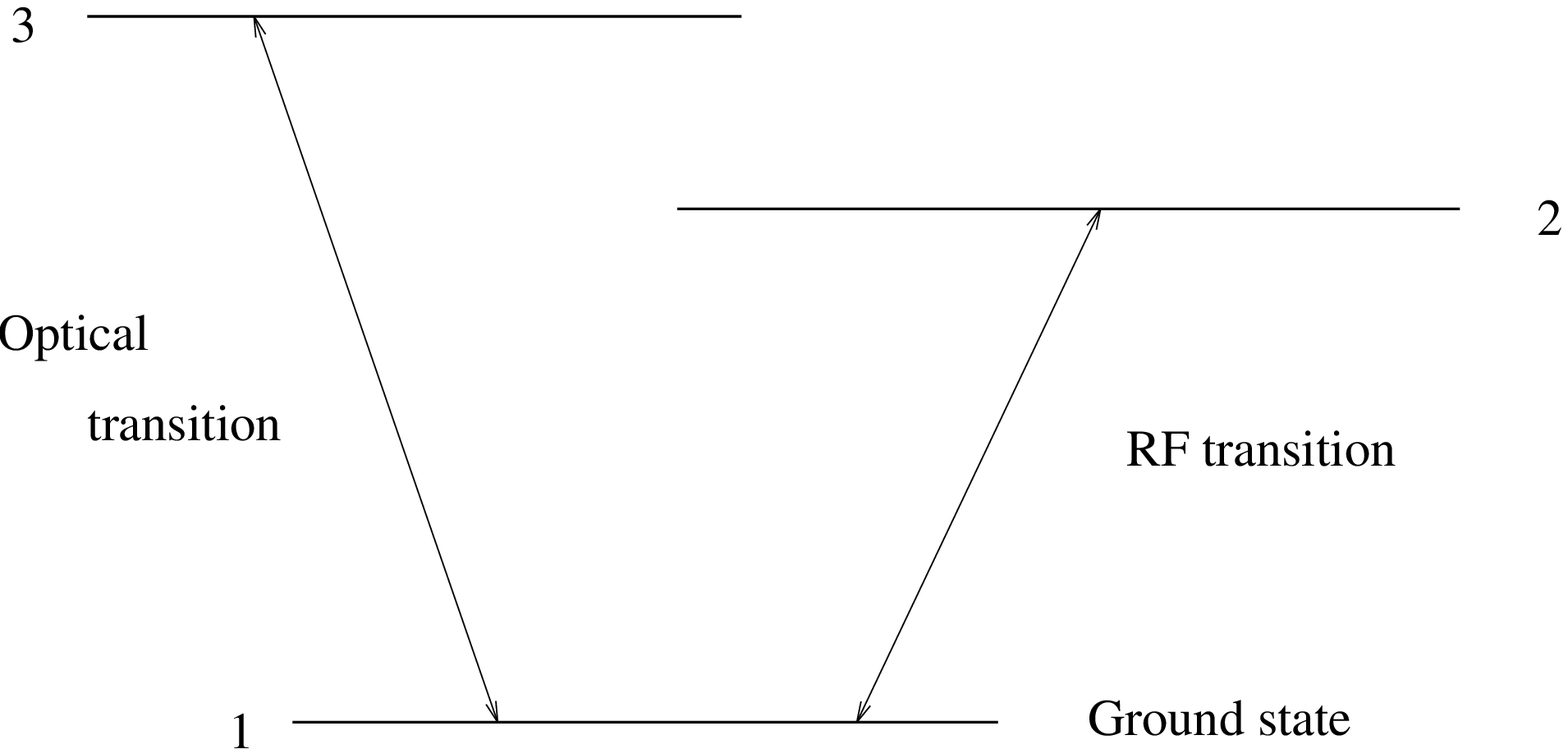}}
{\raggedright Fig. 1 {\it Energy-level diagram for the QZE experiment [4]}}
\medskip
Level 3 can decay only to level 1. The measurement is carried out by
driving the $1\to 3$ transition with a short optical pulse and observing
the presence or absence of spontaneously emitted photons from level 3
corresponding to the atom being projected to the level 1 or 2. A `freezing'
of population in one level as a result of continuous measurements was
observed. By using the postulate of projection or reduction of the wave
function Etano et al have shown that in the limit of infinitely frequent
measurements, the probablity of one of the levels being populated goes
to unity [4]. Frerichs and Schenzle [5] have shown that the outcome of
the experiment can be explained by looking at the optical bloch equations
for the three-level system without appealing to the projection postulate
or the `wave function collapse'.

We propose that the measurement here can be explained by the `environment
induced decoherence' theory [6] which is based on the understanding that
during the measurement process the system is not isolated but coupled to
an external environment, which leads to a decoherence in the reduced
density matrix of the system, driving it to a diagonal form. The crucial
point here is that the collapse does not take place instantaneously but
over a characteristic time scale, the `decoherence time'. Let us consider
the phenomenon of spontaneous emission (SE). SE decay rates emerge naturally
when a completely quantized field treatment including the coupling to the
field vacuum modes is considered in the Weisskopf-Wigner theory [7]. In
the QZE experiment of Etano et al [4] the two-level system (levels 1 and 2)
constitutes the 'system', the level 3 is the `apparatus' and the collection
of vacuum modes coupling to level 3 is the `environment' (Fig. 2).
\epsfysize=1.0in
\centerline{\epsfbox{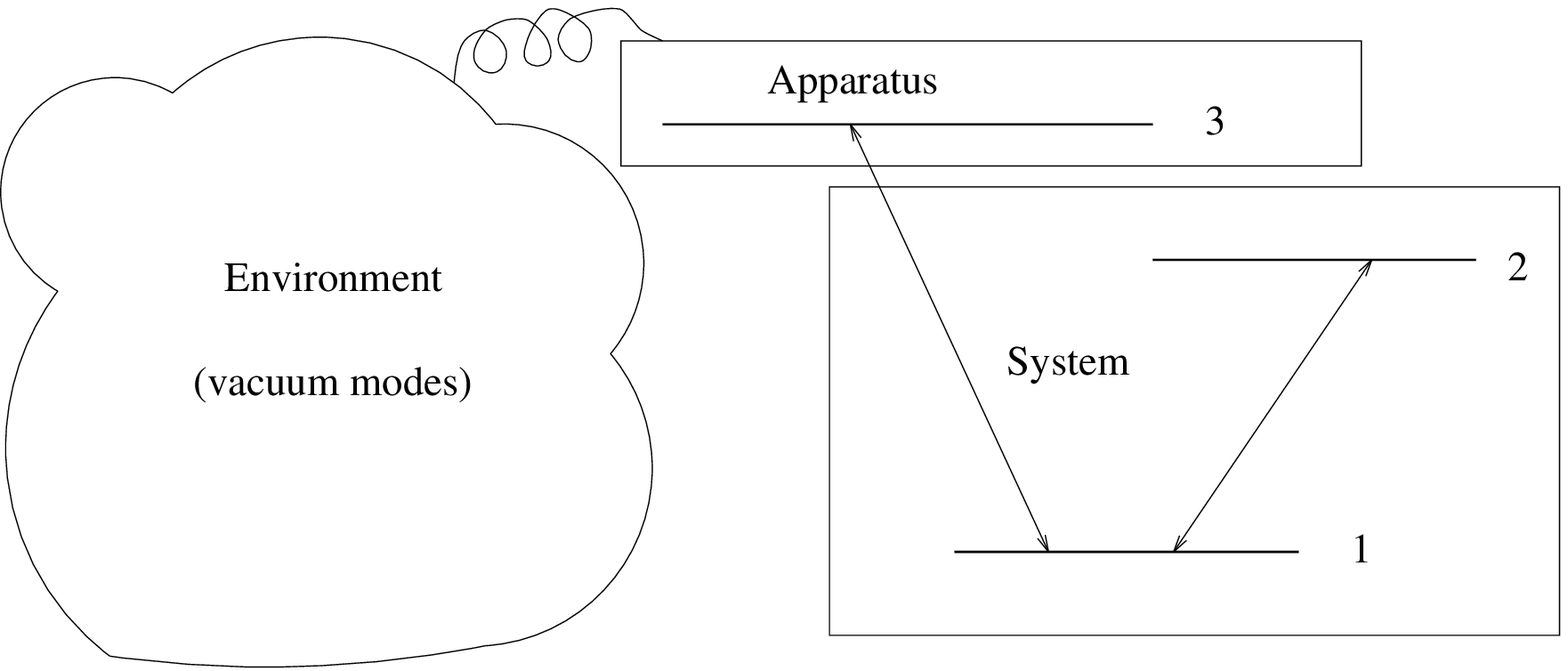}}
{\raggedright Fig. 1 {\it Schematic diagram of the model [8]}}

\medskip
The system-apparatus interaction is through the short pulses that connect
level 1 and 3. The equation for the reduced `system-apparatus' composite
after tracing over the environment variables are the optical Bloch equations
considered by Frerichs and Schenzle [5] which successfully explain the
QZE. It is obvious that the decoherence time for each measurement is the
SE lifetime. It is interesting to note that it sets a fundamental limit
on the requirement of `continous measurements' for QZE since the
measurements cannot be spaced closer than the SE lifetime of the third
level. Interestingly this is the time-energy uncertainty relation that
Ghirardi et al [2] argued about.

To summarize, we have shown [8] how the time-energy uncertainty relation
emerges naturally as a fundamental limit on achieving `continuous
measurements' as required in the QZE when we analyse the QZE problem
using the environment induced decoherence approach. This is in agreement
with the studies of Ghirardi et al [2] and the predictions of Nakazato
et al [3].

\end{document}